%% file: Main.tex
\documentclass[letterpaper,twocolumn,10pt]{article}
\usepackage{flushend}

\usepackage{courier}
\usepackage{enumitem}
\usepackage{usenix,epsfig,endnotes}
\usepackage{filecontents}
\usepackage{tikz}
\usepackage{amssymb}
\usepackage{amsmath}
\usepackage{graphicx}
\usepackage{xurl}
\usepackage[usestackEOL]{stackengine}
\usepackage{seqsplit}
\usepackage{todonotes}
\usepackage{rotating}
\usepackage[russian,english]{babel}
\usepackage{array}

\strutlongstacks{T}

\begin{document}
\title{Entanglement: \\ Cybercrime Connections of an Internet Marketing Forum Population}

\author{%
\begin{tabular}{c} Masarah Paquet-Clouston \\ GoSecure and Université de Montréal \\ m.paquet-clouston@umontreal.ca \\ \\
Sebastián García \\ Czech Technical University \\ sebastian.garcia@agents.fel.cvut.cz \end{tabular} \and
\begin{tabular}{c} Serge-Olivier Paquette \\ SecureWorks \\ spaquette@secureworks.com \\ \\
María José Erquiaga \\ Cisco Systems \\ merquiag@cisco.com \\
\end{tabular} }

\maketitle



\begin{abstract}
Many activities related to cybercrime operations do not require much secrecy, such as developing websites or translating texts. This research provides indications that many users of a popular public internet marketing forum have connections to cybercrime. It does so by investigating the involvement in cybercrime of a population of users interested in internet marketing, both at a micro and macro scale. The research starts with a case study of three users confirmed to be involved in cybercrime and their use of the public forum where users share information about online advertising. It provides a first glimpse that some business with cybercrime connection is being conducted in the clear. The study then pans out to investigate the forum population's ties with cybercrime by finding crossover users, who are users from the public forum who also comment on cybercrime forums. The cybercrime forums on which they discuss are analyzed and crossover users' strength of participation is reported. Also, to assess if they represent a sub-group of the forum population, their posting behavior on the public forum is compared with that of non-crossover users. This blend of analyses shows that (i) a minimum of 7.2\% of the public forum population are crossover users that have ties with cybercrime forums; (ii) their participation in cybercrime forums is limited; and (iii) their posting behavior is relatively indistinguishable from that of non-crossover users. This is the first study to formally quantify how users of an internet marketing public forum, a space for informal exchanges, have ties to cybercrime activities. 
We conclude that crossover users are a substantial part of the population in the public forum, and, even though they have thus far been overlooked, their aggregated effect in the ecosystem must be considered. This study opens new research questions on cybercrime participation that should consider online spaces beyond their cybercrime branding.
\end{abstract}

\section{Introduction}
\label{sec:introduction} 
There is a large spectrum of information technology (IT) tasks surrounding cybercrime that appear legal, like developing websites or redirecting traffic. For such IT tasks, “the criminal character does not have to be clearly visible to the person concerned, or it can be denied afterward”~\cite{leukfeldt2020}[p.6]. Consequently, the \textit{neutrality} of IT~\cite{leukfeldt2020, bijlenga2018}  allows individuals to conduct parts of their cybercrime operations in plain sight. 

This study uncovers and explores the involvement in cybercrime of a population of users interested in internet marketing. It starts with a case study of three actors known to be involved in cybercrime through helping the spread of a banking Trojan botnet. Using a machine learning technique and a content analysis, we assessed the interactions of these three actors as well as their relationships with other users in a public forum. This \textbf{public forum} gathers individuals discussing internet marketing and informally exchanging products and services related to their business.  

The focus of the research then pans out to investigate the forum population's connections to cybercrime by finding crossover users. Crossover users are individuals from the public forum who also commented on cybercrime forums. Three assessments are conducted. First, the population of crossover users is estimated with username matching. Second, we explore the types of cybercrime forums on which crossover users discuss, and their level of participation on these forums. Third, through a series of statistical tests, we evaluate whether the commenting patterns of crossover users, on the public forum, differ from those of non-crossover users. From this series of analyses, we conclude that:  
\begin{itemize}
\itemsep0.5em 
   \item The actors in the case study use the public forum to develop their internet marketing business, a business which has verified connections to cybercrime. 
   \item There is a minimum of 7\% of crossover users in the public forum population.
   \item Cybercrime forums on which crossover users discuss are diverse, from hacking to money laundering and blackhat SEO.
   \item The participation of crossover users in cybercrime forums is limited.
   \item When considering their posting behaviors, crossover users are relatively indistinguishable from other users in the public forum.  
\end{itemize}

This research is the first to explore how users of a public forum on internet marketing have ties to cybercrime. It opens new research avenues on cybercrime participation, avenues that should consider online forums beyond their cybercrime branding, especially given the neutrality of IT tasks~\cite{leukfeldt2020, bijlenga2018}. 

Moreover, the public forum hosts a market where products and services related to internet marketing are exchanged informally. This market recalls traditional informal markets where the product or service is not necessarily illegal; it is rather the means by which it is produced or distributed that is illegal~\cite{ojo2013, hallerportes2005, castellportes1989}. Informal markets are known to be attractive settings for criminal groups to operate in due to their lack of regulations~\cite{ojo2013, mcelwee2011theorising}. This known attractiveness coupled with the findings of this study point towards the need to further investigate how online informal settings can be leveraged for cybercrime operations. 

The paper is divided as follows. Section~\ref{sec:literature-review} presents the literature review. Then, the data, methods and results are presented in Section~\ref{sec:case-study} for the case study (micro scale) and in Section~\ref{sec:crossover-users-section} for the public forum population (macro scale). Section~\ref{sec:discussion} and Section~\ref{sec:limits} present the discussion and the study limits respectively. A short conclusion is provided in Section~\ref{sec:conclusion} 

\label{sec:literature-review}

To frame the results of this study, Section~\ref{sec:beyond} presents previous work on the cybercrime industry and the IT tasks surrounding cybercrime operations. Then, a short overview of informal markets and their online counterparts is provided in Section~\ref{sec:informal}  . 

\subsection{\textbf{Beyond the Cybercrime Underground}}
\label{sec:beyond}

Understanding the organization of the cybercrime industry has been a topic of interest in computer security and criminology in the past decades~\cite{anderson2019, Wegberg2018, afroz2011, collier2020, manky2013cybercrime, huang2018systematically, lusthaus2018industry}. A key feature of the industry is specialization: one can specialize in a specific task, such as monetizing credit cards, and outsource the remaining tasks to other actors in the industry~\cite{Thomas2015,lusthaus2018industry, manky2013cybercrime, huang2018systematically,hutchings2015crime,Wegberg2018}. Such specialization reduces the costs of cybercrime through increased productivity and profitability~\cite{moore2009}. 

Two recent studies~\cite{Wegberg2018,akyazi2021measuring} investigated specifically "as-a-service" advertisements in underground forums. One~\cite{Wegberg2018} focused on eight online anonymous markets (also known as darknet markets~\cite{broseus2017geographical} or cryptomarkets~\cite{martin2014}) over six years and the other~\cite{akyazi2021measuring} on the well-known underground forum named "HackerForums" over 11 years. In both cases, despite clear evidence of specialization in the industry~\cite{Thomas2015,lusthaus2018industry, manky2013cybercrime, huang2018systematically,hutchings2015crime,Wegberg2018}, the two studies showed that the number of specialized "as-a-service" listings advertised in underground forums was limited. From these findings, the authors of~\cite{Wegberg2018} hypothesized that outsourcing critical parts of the cybercrime value chain  may be difficult. On the other hand, such "as-a-service" offerings may be limited because a great number of tasks related to cybercrime operations require IT expertise, but not necessarily secrecy.  

Several studies reported criminal groups actively seeking such expertise~\cite{leukfeldt2017a, leukfeldt2017b, leukfeldt2017c, leukfeldt2017d,bijlenga2018}. For example, when studying networks involved in banking theft, ~\cite{leukfeldt2017a, leukfeldt2017b, leukfeldt2017c, leukfeldt2017d} reported core members of criminal groups recruiting individuals to develop websites (programmers) or translate texts. Some of these tasks were not criminal, but their use was. 

Bijlenga and Kleemans (2018)~\cite{bijlenga2018} also found that individuals and organizations with IT expertise were actively leveraged by criminal groups. They studied five Dutch criminal investigations where expertise in the IT sector was sought by individuals involved in criminal activities. In three of the five cases, the basis of the collaborations was a legal business relationship. The authors mentioned that such a relationship was possible because the criminal nature of the tasks was not always obvious; the good or service provided was legal, while its usage was not. 

When discussing criminal groups seeking IT expertise,~\cite{leukfeldt2020} stated that business collaborations can be established without the contractor or seller knowing that the product or service provided will be used for criminal means. The authors argued, that due to the neutrality of IT, “the criminal character does not have to be clearly visible to the person concerned or it can be denied afterward”~\cite{leukfeldt2020}[p.6]. This neutrality creates a blurry frontier between legal and criminal IT tasks, allowing individuals to recruit beyond underground settings. Plus, beyond these settings, there exist informal markets that represent interesting spaces to find business partners.  

\subsection{\textbf{The Middle-Ground: Informal Markets}}
\label{sec:informal}
The concept of \textit{informality} is a broad and multifaceted concept tackled by scholars from various disciplines, including economics, sociology, and criminology. In general, informal markets are associated with the reverse side of the official economy: the unregulated or unregistered economic activities~\cite{ponsaers2008}. In such markets, the product or the service exchanged is not necessarily illegal; it is rather the means by which it is produced and distributed that is illegal~\cite{castellportes1989, ojo2013}. For example, developing a website for commercial purposes and not declaring the profit associated with it represents an informal economic activity. On the other hand, using the developed website for a cybercrime operation that steals banking credentials represents a criminal activity. 

In traditional settings, informal markets are known to be attractive for criminal groups due to their lack of regulations~\cite{sabet2015informality, shapland2004informal,walle2008matrix}. For example, informal financial markets can easily be leveraged for money laundering~\cite{walle2008matrix}. Also, a study of 30 informal entrepreneurs in the UK~\cite{ojo2013} illustrated that informal market participants are ready to embark on criminal business opportunities when the prospects for profits are high and the likelihood of getting caught is low.

Portes and Haller (2005)~\cite{hallerportes2005} define three aims of informal economies for market participants: survival, dependent exploitation (such as decreased labor costs) and growth, the latter including capital accumulation, solidarity and flexibility [p.405-6]. Thus, considering the latter, informal economies are not solely destructive; they also provide jobs to otherwise unemployed individuals, lower costs for products and services and foster innovation~\cite{ojo2013, hallerportes2005}. Moreover, ~\cite{sabet2015informality} studied the relationship between informal workers and crime and found results different from those of the study on 30 informal entrepreneurs in the UK~\cite{ojo2013}. Sabet (2015)'s study illustrated that although, in theory, informal settings offer opportunities for criminal groups, in practice, those who operate in informal settings tend to avoid being involved in criminal activities when possible~\cite{sabet2015informality}. 

Online settings are known to foster unregulated and unregistered economic activities~\cite{dobson2015, murray2007, ponsaers2008, rangaswamy2019}. Online informal markets, just like the offline counterparts, create an environment that is auspicious for criminal activities. For example, freelancer platforms are known to host informal economic activities~\cite{schmidt2017, drahokoupil2017, drahokoupil2016}) and two of these platforms have already been associated with cybercrime activities. Farooqi et al. (2017)~\cite{farooqi2017} tagged the platform SEOClerk as a “blackhat marketplace” and~\cite{garg2013} considered the platform Freelancer as a hub for criminal activities. The latter assumption was based on the results of~\cite{motoyama2011}, a study that investigated the on-demand platform Freelancer and concluded that 66\% of the jobs posted were legitimate, meaning that about 33\% of the remaining tasks were likely related to illegal activities, included thwarting security mechanisms or sending spam.

This study uncovers and explores the involvement in cybercrime of a population of users interested in internet marketing, both at a micro and macro scale. What gathers this population is a legal public forum on which users discuss internet marketing and informally exchange products and services related to their business. The case study is presented first (micro), followed by the analysis of crossover users (macro).

\section{Case Study}\label{sec:case-study}
The study starts with a case study of three actors involved in cybercrime and their use of the public forum. We position them with respect to other forum users by using machine learning (the forum map) and conduct a content analysis on their forum interactions. These analyses allow us to better understand the role of the public forum for these actors, as explained below. In this section, the context of the case study is first presented, including an introduction of the public forum and how we gathered its publicly available data. The data and methods are then presented, followed by the results of the case study. 

\subsection{\textbf{Context}}
The case study builds on a previous research~\cite{paquet2021role} that investigated the private conversations of individuals known to be involved in cybercrime activities. The \textit{three main actors} in the private conversations (named Actor 1, Actor 2 and Actor 3 in this study)\footnote{They represent the Main Entrepreneur, Website Master 1 and Developer 1 in~\cite{paquet2021role}.} developed websites advertised as libraries for “cracked” or “modded” Android applications (APKs). Modded APKs are modified versions of originals usually providing either better functionalities or unlocking paid features. 

Between 2017 and 2018, when the private conversations took place, the APKs available on their website were malicious and related to the Geost botnet. The Geost botnet was an Android banking Trojan botnet that infected nearly 800,000 Russian phones and had access to million of Euros~\cite{Garcia2019VB,Garcia2019}. When visiting the websites of the three actors, visitors thought they were downloading modded or cracked APKs, while what they were actually downloading were banking Trojans. The three actors were paid for every malicious APK that was successfully installed through their websites. They acted as affiliates in what appeared to be a black market pay-per-install (PPI) program related to the Geost botnet~\cite{Garcia2019VB,Garcia2019}. 

The previous research~\cite{paquet2021role} showed the difficulties that these actors faced daily. They were amateurs trying to monetize their websites through any means necessary, participating in various monetization programs, both legal and illegal. The three actors also discussed on a public forum, the focal point of this study. We associated the actors in the private conversations with their public forum usernames because (i) they used the same usernames; and (ii) they pasted in the private conversations links to their public interventions, such as: \textit{“ordered texts [link to comment on the public platform]”}.

\subsubsection{\textbf{Introducing the Public Forum}}
This forum is a Russian and English-speaking forum dedicated to internet marketing.  It was created in early 2000 and, as of 2021, reported over 400,000 registered members and 14,000,000 comments. It advertises itself as a \textit{“website allowing users to discuss issues related to creating and promoting websites on the internet [...]. The forum brings together experts in all areas of online advertising and allows you to receive both free knowledge and find mutually beneficial contacts and partners”}. Topics of discussion are divided into categories which range from search engine result optimization to monetizing sites or hiring web masters. Although the forum is not an official matchmaker for demand and supply of products and services related to internet marketing, many users leverage it as an advertisement space. Hence, just as the actor above did when he ordered texts for his websites (he did so by answering a post in which another user offered such a service), many users conduct business deals through the forum. In this study, the public forum is conceptualized as a space where informal exchanges of products and services related to internet marketing take place.


\subsubsection{\textbf{Data Access}}
Note that all information on the public forum was extracted through an academic access to the Flare Systems database. Flare Systems \cite{FlareSystems} is a Montreal-based company that has developed a digital risk protection and cyber threat intelligence platform. Since 2016, it has been monitoring this public forum along with over 100 others. To make sure that Flare Systems' database is representative, we selected fifty random actors and compared the number of comments found on the database with the number of comments found on the forum from 2012 to 2020. Flare Systems had, on average, 93\% (std=0.13) of the total number of comments published on the forum, illustrating a sufficiently accurate coverage for our research.

\subsection{\textbf{Creating a Forum Population Map}}
To first gain a comprehensive perspective on the public forum population, we positioned the three actors in the public forum relative to others, based on their posting behavior in each of the forum's categories. This was possible by generating a forum map (i.e., a visual aid). The forum map is a two dimensional representation of the forum population where each individual is graphically put in relationship with each other. The dataset created for this analysis is presented below.



\subsubsection{\textbf{Dataset of the Forum Population}}
Since the public forum has been active for nearly 20 years, we selected the timeframe of the private discussions, 2017 and 2018, to be the study period for the forum map dataset. Selecting these two years allows us to stay as close as possible to the context of the private discussions when mapping the forum population in relation with the three actors. 

More precisely, using the Flare Systems API, we extracted all comments posted on the public forum between 2017 and 2018. For each comment, the extracted features were: the comment’s identification number, text, timestamp, the name of the actor who wrote it, the title of the thread, and the thread’s identification number. The final dataset included \textbf{685,815} comments, \textbf{34,706} threads and \textbf{23,348} users\footnote{One username was \selectlanguage{russian} “Этот пользователь удален” \selectlanguage{english} which means “This user has been deleted”. Consequently, we removed all comments related to this specific username.}

To be able to map users based on their posting behavior in each category of the public forum, we had to extract the thread's category. Consequently, we crawled the public forum over several days (so as to ensure that the website's server would not experience disruption from our research activity) using the thread identification number. A total of nine categories and 80 subcategories was found. Each category has a specific set of rules enforced by the public forum administrators. Table~\ref{tables:Categories} shows the nine categories and a sample of their \textit{subcategories}, along with the percentage of comments posted in each category in 2017 and 2018 combined. \textbf{Categories} ranged from \textit{Search Engine} to \textit{Monetizing Websites} to \textit{Hiring Webmasters}. As shown in  Table~\ref{tables:Categories}, the category \textit{About Monetizing Sites} was the most popular one, representing 20\% of the comments, followed by \textit{Not About Work} with 17\% and \textit{Site Building} with 16\%.

\begin{table*}[h]
\small
\caption{Summary of categories and subcategories in the public forum, with the percentage of total comments on them between 2017 and 2018.}
\begin{center}
\label{tables:Categories}
\input{tables/Categories}
\end{center}
\end{table*}

\subsubsection{\textbf{Descriptive Statistics}}
Table~\ref{tables:descriptive_stratistics_dataset_2017_2018} presents the descriptive statistics of the forum population dataset. In this dataset, users commented, on average, 30 times (std = 151) on 11 threads (std=50) and two categories (std = 2). At least 50\% of users commented fewer than four times, illustrating that user participation is unequal, with most users exhibiting a low rate of participation. This distribution of comments reflects the participation inequality rule found in online communities and highlighted by several  scholars~\cite{haklay2016, paquet2018, sun2014, mooney2012, lund2011, vanmierlo2014}. 

\paragraph{Top Poster Dataset.} Due to non-participating users blurring the visual representation, using the entire dataset to map the forum population was inconvenient. When investigating the distribution of comments per user, we noticed a slight breakdown at 10 comments, with about 70\% of users posting fewer than ten comments and 30\% posting more than ten comments. Consequently, we used this slight breakdown to reduce the noise induced by a mass of sporadic users and created a subset of the dataset with \textit{Top Posters}: those who posted at least 10 times in 2017 and 2018. Descriptive statistics of the \textit{Top Poster} dataset are presented in Table~\ref{tables:descriptive_stratistics_dataset_2017_2018} as well. In this dataset, users commented, on average, 92 times (std = 267) on 34 threads (std=88) and four categories (std = 2). At least 50\% commented fewer than 27 times. We used this {Top Poster} dataset to create the public forum map presented below.

\begin{table}[h]
\caption{Descriptive statistics for the forum population dataset and the \textit{Top Poster} dataset}
\begin{center}
\label{tables:descriptive_stratistics_dataset_2017_2018}
\input{tables/descriptive_stratistics_dataset_2017_2018}
\end{center}
\end{table}

\subsubsection{\textbf{Uniform Manifold Approximation and Projection for Dimension Reduction (UMAP)}}
To project the data points of users, based on their posting behavior in each of the forum's categories, into a comprehensible representation, the Uniform Manifold Approximation and Projection for Dimension Reduction (UMAP)~\cite{mcinnes2018umap} was used. UMAP is a mathematically robust and efficient method to project high dimensional data into lower dimensions while preserving the underlying structure both at the local and global scales. Due to its ability to embed complex structural relationships into much more comprehensible representations, UMAP has recently been used to map highly complex phenomena like cellular biology~\cite{cao2019single,packer2019lineage} and multi-scale population structure~\cite{diaz2019revealing}. 

The data of the users was projected as points from $\mathbb{R}^9$ (number of forum categories) to the visual plane $\mathbb{R}^2$. Each dimension in $\mathbb{R}^9$ represents a feature, which corresponds to the number of comments made by that user in one of the nine categories over the span of the two years. Note that the two coordinates in the $\mathbb{R}^2$ projection do not have a semantic meaning. However, the distance between points in $\mathbb{R}^2$ respects the underlying manifold structure in the source $\mathbb{R}^9$ dimensions as much as possible. 

UMAP has two important parameters: the minimal number of neighbors for each point in the original dimension and a distance measure. Both parameters of UMAP were tuned as a result of an exploratory phase and the distance measure needed to be justified according to the nature of the data and the inquiry. The final parameters used for the UMAP transformation, chosen so that it resulted in the clearest segmentation of users, were \textit{n\_neighbors}=100 along with the Euclidean distance. Euclidean distance  is a generalization in \textit{n} dimensions of the natural notion of distance between points in a plane ---a straight line. 

We added the three actors to the forum map, providing an insightful overview of their position in the forum population (as presented in Section~\ref{sec:results-case}). Additionally, for the case study, we conducted a content analysis on the three actors' public comments, giving us a qualitative understanding of their use of the forum.

\subsection{\textbf{Analyzing Actors' Comments}}
To gain a deeper understanding of the actors' use of the public forum, we conducted a content analysis of their publicly posted comments. To do so, we gathered all available information about them from the public forum using the Flare Systems' database, as explained below. 

\subsubsection{\textbf{Dataset of Actors' Comments}}
Using Flare Systems' database, we extracted all comments\footnote{Note that in this study, the term \textit{comment} is used to describe a piece of text that was sent, or \textit{posted}, as part of an interaction. Others may refer to them as \textit{posts}, \textit{interactions} or \textit{interventions}.} for each of the three actors, including the comment's timestamp, identification number, content, related actor's identification number, title of the thread, and thread identification number. Table~\ref{tables:three_actors_public_comments} presents the number of comments and threads found on the public forum for all three actors, as well as their first and last year of activity. All comments were written in Russian and translated using the \textit{Googletrans} library~\cite{GoogleTrans}. To have a global overview of actors' use of the forum, we decided to analyze all their comments, regardless of the time they were posted. This allowed us to assess whether the actors ever mentioned cybercrime participation in the public forum.

\begin{table}[h]
\caption{Three Actors' Participation in the Public Forum}
\begin{center}
\label{tables:three_actors_public_comments}
\input{tables/three_actors_public_comments}
\end{center}
\end{table}

\subsubsection{\textbf{Content Analysis}}

We analyzed the comments using a content analysis with the research question: \textit{how do the actors interact in the public forum?} Content analysis aims at systematically uncovering the use of certain words, themes and semantics. More precisely, we developed an initial coding scheme with a small subset of 215 comments. Then, one coder followed that scheme to classify the three actors' comments which summed to 1,990 comments from 2009 to 2020. In situations of uncertainty, the coder consulted the rest of the team, who evaluated the comments and agreed on a theme. 

To contextualize comments, and because the automated translation was sometimes inadequate, the coder often searched the comments on the public forum and went through the related thread using web browser translation (that showed better translation results). In extreme cases, such as the use of slang words, the coder consulted a native Russian-speaking collaborator to understand the meaning of the comment. The meaning of each theme evolved throughout the analysis, as the coder included more comments in each of them. Finally, the coder also created memos on each actor to complement the themes. At the end of the analysis, four large themes encompassed how the actors interacted on the public forum: as a \textbf{buyer}, as a \textbf{seller}, as a \textbf{forum participant} or as a \textbf{tool user}. Each of these is described below.

\textbf{\textit{As a Buyer}}
The theme \textit{as a buyer} included comments aimed at purchasing or commenting upon the quality of a service or a product, including providing a positive, neutral or negative feedback, asking for a price or asking to contact a seller for further information. 

\textbf{\textit{As a Seller}} 
The theme \textit{as a seller} included comments aimed at offering a service or a product in the public forum, thanking a buyer when receiving feedback or providing customer services. 

\textbf{\textit{As a Tool User}} 
The theme \textit{as a tool user} included comments aimed at giving an opinion about a tool, helping others to solve similar issues or sharing experience with a tool (such as website promotions tools or traffic monitoring tools).

\textbf{\textit{As a Forum Participant}} 
The theme \textit{as a forum participant} included comments that were more related to participating in the public forum in general, such as asking general questions or giving general advice (not specific to a tool) as well as sharing information on various topics. 

\subsubsection{\textbf{Ethical Considerations}}
The case study presented above has been approved by the Simon Fraser University ethics department *(study number 2020s0121) (former institution of the lead author) and University of Montreal (study number CERSC-2021-131-D) under minimal risks, which required asking for a \textit{waiver of consent} in line with Article 5.5A of the TCPS2. To ensure participant’s confidentiality and privacy, we do not use the real pseudonyms of the actors. 

There are ethical issues regarding the research that need to be acknowledged~\cite{thomas2017ethical}. In terms of potential harms, studying the cybercrime connections of a public population can lead to  wrongly labeling individuals as cybercriminals. It can also result in profiling and marginalizing forum users, while also shifting law enforcement focus onto them. We try to avoid creating these harms by taking a nuanced approach when interpreting the results. In return, the research leads to better understanding the context that may lead a mass of users to have connections with cybercrime. It uncovers and discusses a largely ignored phenomenon: how the neutrality of IT can lead to cybercrime participation in legitimate forums. Finally, the interpretation of the results leads to policy opportunities that can \textit{prevent} cybercrime participation.

\subsection{\textbf{Results of the Case Study}}
\label{sec:results-case}
The results of the case study are presented below, including the forum map and the three actors' positions in it as well as the findings from the content analysis. 


\subsubsection{\textbf{The Map of the Forum Population}}
\label{sec:mapping-forum-population}
The resulting two dimensional representation of the public forum users is shown in Figure 1 and is called \textit{the map} in this work. In \textit{the map}, each point represents a user, and points that are close together are users with similar posting profiles. The (x,y) coordinates are not directly interpretable but the relationships between points are important. Figure 1 shows a set of arm-shaped groups that stretches from the center to the outside. The shapes of the groups are very informative, as groups that take the shape of long and narrow arms represent users that comment mostly in one category, with the most active users at the outer ends of the arms. 

We investigated each of the arms in the map to find the dominating public forum category on that arm. These groups define more or less tight communities, some with very active users, commenting several thousand times over the span of two years, as in the \textit{Site Building} category. On the other hand, the arms that are wider or closer together can be seen as groups that blend more or less with other ones, as in the \textit{Not About Work} and \textit{Communication of Professionals} categories. Also, some categories like \textit{Work and Services for Webmasters} and \textit{Exchange and Sales} are close to each other, but do not contain users with extreme posting behavior.

\begin{figure}
    \centering
    \includegraphics[width=0.4\textwidth]{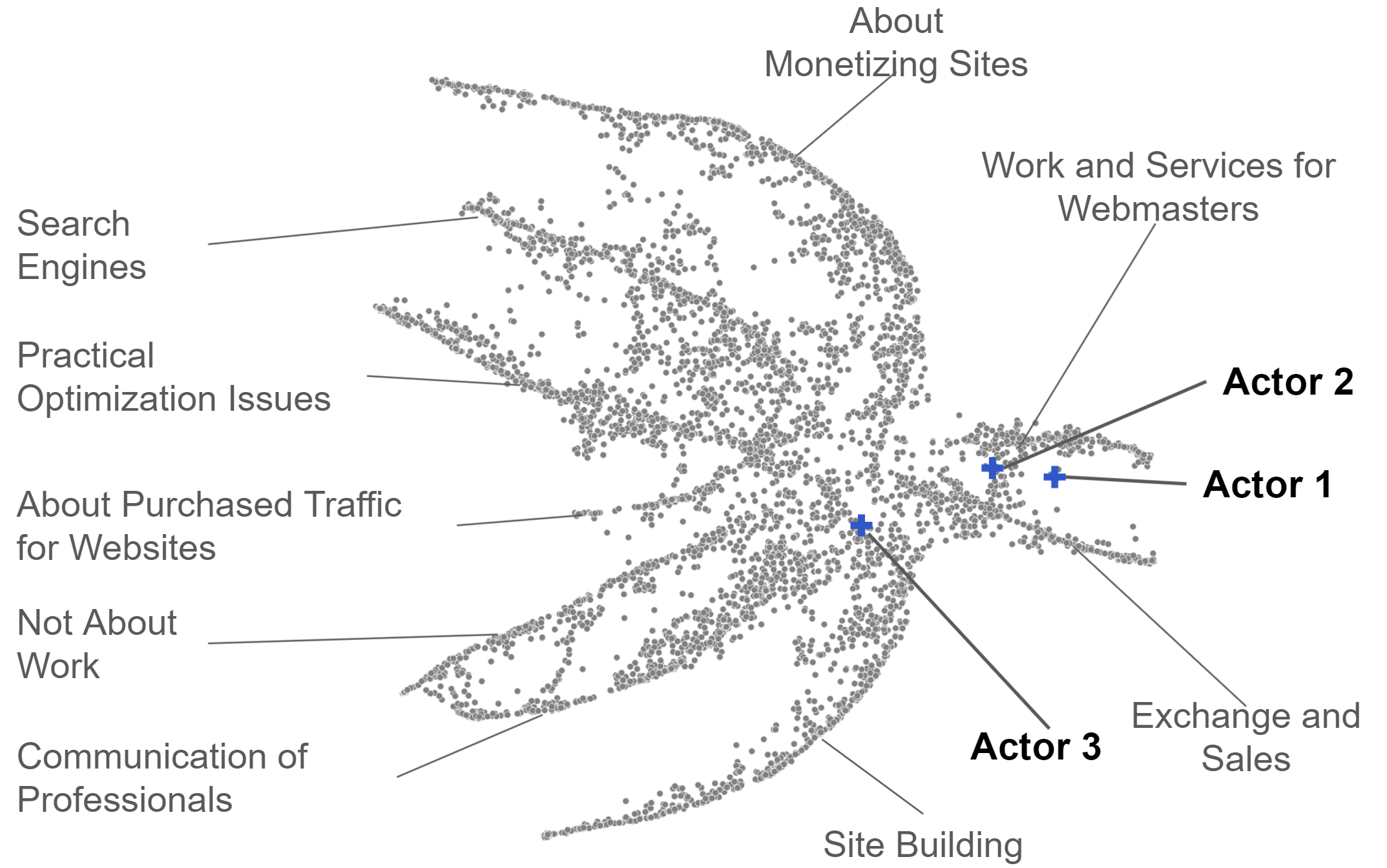}
    \caption{The Map of the forum with the 2017-2018 Top Poster dataset}
    \label{fig:bird_categories_anonymized}   
    \centering
\end{figure}

\subsubsection{\textbf{Positioning the Three Actors}}
We added the three actors to the map in Figure 1. Their location is quite informative: they are closer to the center than to any arm's end. The shapes of the groups which Actors 1 and 2 are close to, \textit{Work and Services for Webmasters} and \textit{Exchange and Sales}, are also interesting. These are much shorter arms and are closer together, indicating that fewer users in these categories were very active contributors to their groups, hinting at a more opportunistic behavior. These categories are also explicitly related to business. Actor 3, on the other hand, does not seem to belong to any group of interest on this platform. 

\subsubsection{\textbf{Typical Users}}
Figure 2 shows the distribution of themes found for each actor based on his comments on the public forum. These themes are discussed below for each actor, along with additional contextual (and valuable) information that was available in the coder's memos. 

\textbf{Actor 1} 
Figure 2 shows that Actor 1 interacted \textit{as a buyer} 62\% of the time, buying a variety of web products, including images, logos, texts, code reviews, security reviews, programs, scripts, and traffic. We noticed an overlap between the actor's comments in the private discussions and his comments in the public forum. For example, in the private discussions, Actor 1 discussed the need to review one of his websites for search engine optimization, this website is known to have hosted more than 17 malicious Geost APKs. In the public forum, within the same timeframe, he asked for an external review of one of his sites (a transaction that was successfully completed, according to the public conversations).  Also, Actor 1 interacted \textit{as a forum participant} 19\% of the time, sporadically helping others on website building matters. The actor also interacted \textit{as a seller} 10\% of the time, offering writing services, ready-to-use APK portals ---known as turnkey websites--- as well as generic templates and website layouts. Lastly, the actor interacted only 9\%  \textit{as a tool user}. 

\textbf{Actor 2} 
The second actor was less active on the public forum, posting a total of 169 comments. As shown in Figure 2, the actor interacted 37\% of the time \textit{as a forum participant}, commenting on topics related to programming and website traffic or recommending websites in general, and 26\% of the time \textit{as a tool user}. He also interacted 21\% of the time \textit{as a seller}, selling ready-to-use APK portals, as well as generic APKs, images, videos or texts. Lastly, about 16\% of the time, Actor 2 acted \textit{as a buyer}, purchasing, for example, systems to monitor sites or texts to fill websites. 

\textbf{Actor 3} 
Actor 3 commented 457 times on the public forum and 76\% of his comments were \textit{as a forum participant}, helping others or asking for help on internet marketing topics. Only 15\% of the time did Actor 3 interact \textit{as a seller}, offering, for example, scripts and parsers. He also interacted in the public forum 9\% of the time \textit{as a buyer}, purchasing tutorials for social media marketing or ready-to-use websites. None of his comments were \textit{as a tool user}. 


\begin{figure}
  \centering
    \includegraphics[width=0.5\textwidth]{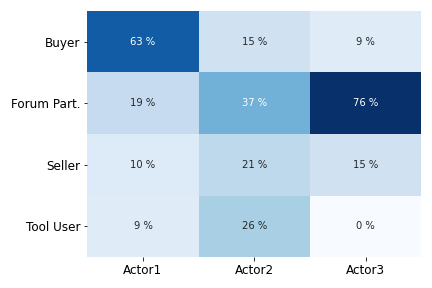}
    \caption{Distribution of Themes per Actor}
    \label{fig:positions}   
  \centering
\end{figure}

In sum, all three actors interacted \textit{as a seller} or \textit{as a buyer} (Actor 3 to a lesser extent) and many of the topics they discussed in the public forum were the same topics discussed in the private conversations. Consequently, the three actors actively used the public forum as source of information, products and services related to their business, which by 2017 and 2018, had ties to cybercrime activities. According to the forum map, Actors 1 and 2 were positioned in more opportunistic groups with users who spoke less, on average.  Actor 3 was positioned near the center map, in no specific groups. There is nothing special about them; they are typical users of the forum. We noticed, as well, that none of the comments studied hinted that these actors were involved in cybercrime activities. 

Given the results of the case study, whether other users in the public forum had connections to cybercrime became a topic of interest. This allowed us to move from an in-depth micro understanding of three actors to a macro assessment of the scale of the problem. In the next section, we  analyze crossover users, or public forum users who also participate in cybercrime forums.

\section{Deep Dive on Crossover Users}
\label{sec:crossover-users-section}

The forum population overlap with cybercrime spaces is assessed by identifying \textbf{crossover users: individuals from the public forum who also discussed on cybercrime forums}. We evaluate the scale of the problem through a series of analyses. Crossover users are first identified and the cybercrime forums on which they discuss are analyzed. Then, their posting behavior on the public forum is compared with non-crossover users. The idea is that, if crossover users form a  subgroup in the public forum by displaying specific posting behaviors, they should be further investigated as a subgroup of the forum population. The data and methods used for these analyses are first presented below, followed by the results.

\subsection{\textbf{Strategy to Identify Crossover Users}}
To identify crossover users, we searched in the Flare Systems database whether some of the usernames found in the public forum also discussed in cybercrime forums during a similar timeframe. This cross-correlation method is based on the idea that users are likely to choose the same username in different forums, a phenomenon that was observed in previous studies~\cite{wang2016,perito2011}.  

We do acknowledge that this approach will inevitably yield false positives, but these two studies~\cite{wang2016,perito2011} suggest that our methods will yield more false negatives (those we missed with the method) than false positives (those we flagged as crossover users when they are not). This inequality leads us to infer a \textit{lower bound} on the number of crossover users.



\subsubsection{\textbf{Filtering to Find Crossover Users}}
For this analysis, we used the same dataset of all users who posted in 2017 and 2018 in the public forum (generated for the UMAP analysis). This dataset included a total of \textbf{685,815} comments, \textbf{34,706} threads and \textbf{23,348} users. From this dataset, we created a list of unique usernames and developed two types of filters: username filters and timeframe filters. Also, only public forum users who commented on forums that had a clear cybercrime branding were identified as crossover users, as explained below.

\paragraph{\textbf{Username Filters.}} 
We filtered the list of unique public forum usernames to keep only  those with at least five characters, removing common usernames, such as “Nick”, “Max,” or “bot”. The general idea is to minimize the chances of cross-correlating generic usernames (due to their popularity or lack of sophistication) and maximize the chances of keeping one-of-a-kind usernames. We developed additional -more conservative- filters, but the results did not change the narrative of the paper. For the sake of concision, only the results of the most liberal filter are presented below, gathering as many users as possible who could be crossover users~\footnote{We developed two more conservative usernames filters: one with a minimum of six characters and another one with a minimum of five characters AND at least an uppercase OR a number OR one special character. Because these filters were stricter, they reduced the number of crossover users identified. We computed all statistical tests with these additional filters and the results were similar. Although they are not presented in this study, they can be found in the PhD thesis~\cite{paquet2021role}}.

\paragraph{\textbf{Timeframe Filters.}} We searched for the filtered five-character usernames in the Flare Systems database to find if some of them commented in other forums monitored by Flare Systems. However, since Flare Systems has visibility on forums that have been active since early 2000, we had to filter out the comments based on the year they were posted. Since the timeframe of the public forum dataset presented above is 2017 and 2018, we heuristically decided to keep only comments posted from 2015 to 2020; effectively extending the time range by 100\% before and after the timeframe of the dataset.

This aimed at minimizing the chances of cross-correlating usernames that might belong to different individuals because of the time difference between the comments posted, while also generating a sufficient dataset for the analysis. We also developed more conservative timeframe filters, but the results, again, did not change the narrative of the paper. For the sake of concision, they are thus not presented below~\footnote{We developed additional conservative timeframe filters: one considered all comments posted on cybercrime forums one year prior to 2017 and one year after 2018, thus from 2016 to 2019. Another one considered only comments posted on the cybercrime forums during the period of study: in 2017 and 2018. Again, we computed all statistical tests with these additional filters and the results were similar. Although they are not presented in this study, they can be found in the PhD thesis~\cite{paquet2021role}}.

\paragraph{\textbf{Selecting Forums with a Cybercrime Branding.}}
If a user with a five-character username commented on a forum (other than the public forum) during the timeframe mentioned above, we extracted the identification number for the comment, its timestamp, and the name of the cybercrime forum on which it was posted. This resulted in identifying 42 forums where crossover users commented. We manually verified whether these 42 forums had a cybercrime branding. Four of the 42 forums were branded as a forum for cybersecurity discussions, and were therefore removed. The remaining 38 forums were openly related to cybercrime activities. 

All in all, the \textbf{crossover user dataset} included all comments posted between 2015 and 2020 in one of the 38 verified cybercrime forums by public forum users who had usernames of at least five characters. We added a binary variable identifying crossover users to the forum population dataset. 

\subsection{\textbf{Investigating Cybercrime Forums}}
We also visited the cybercrime forums to codify their main branding, such as cracking or money laundering. The code was determined based on the website's official description and/or the main topics discussed in the front page. Sometimes, these forums were down or required registration. In such cases, to find their general branding, we extracted information about the forum in security reports and blogs or in the Flare Systems database. The main branding identified sometimes overlapped (e.g., hacking and blackhat SEO forums host similar discussion topics), and in such cases, the most obvious one was kept. Also, since the branding of some forums was not well-defined, we created a \textit{catchall} code named: \textit{discussions, sales and questions on various content related to cybercrime}. The idea is to provide a general picture of the types of cybercrime forums on which crossover users discussed.

We also coded whether the cybercrime forums were hosted on the \textit{clearnet}, meaning that they could be visited via a modern web browser, such as Google, or hosted on The Onion Router (Tor), known as the \textit{darknet}. Tor is an anonymous communication protocol developed by a network of volunteers that allows users to browse the internet anonymously~\cite{Tor}. The anonymous protocol also hosts websites, known as onion services, that offer anonymity to both website owners and visitors. These onion services are often associated with the \textit{darknet}~\cite{fidalgo2019classifying,broadhurst2020availability,owen2015tor}, a loosely defined concept that encompasses networks that are not accessible via modern web browsers and offer anonymity to their users, such as I2P, Freenet, Tor, and ZeroNet~\cite{hu2020traffic}. Content is more likely to be related to criminal activities when hosted on these technologies due to the anonymity provided~\cite{owen2015tor}. 

\subsection{\textbf{Distinguishing Crossover Users in the Public Forum}}
Finally, to compare crossover users with non-crossover users on the public forum, we developed a series of posting behavior indicators. We then compared crossover users with non-crossover users based on these indicators through a series of non-parametric tests. We computed the analysis  twice, once on the \textit{Top Poster} dataset, to remove the potential effect of a mass of non-participating users, and once on the entire forum population dataset. 

\subsubsection{\textbf{Posting Behavior Indicators}}
We developed three indicators with 13 sub-indicators to measure posting behaviors. They are presented below.

\textit{\textbf{Activity Rate}} The first indicator quantifies the extent to which a user is active on the public forum. It includes two sub-indicators: (1) \textit{N. Posts} which is the sum of all comments made by each user in 2017 and 2018, and (2) \textit{N. days active} which is the number of days each user was active over the two-year period (meaning the number of days the user posted at least once). 

\textit{\textbf{Diversification}} The second indicator quantifies the extent to which a user is diversified in the  public forum. It includes two sub-indicators: (1) \textit{N. cat}: the number of categories in which the user commented (categories are listed in Table~\ref{tables:Categories}), and (2) \textit{N. sub-cat}: the number of subcategories in which the user commented (a sample of subcategories is listed in Table~\ref{tables:Categories}). 

\textit{\textbf{Topics Discussed}} The third indicators measures the extent to which a user is active in a specific category of the public forum. This indicator thus includes nine sub-indicators, one per category. Each sub-indicator includes the total number of comments posted by a user in one of the nine categories. 

Descriptive statistics, for each sub-indicator, are shown in Table~\ref{tables:descriptive_statistics_indicators} for the \textit{Top Poster} dataset. The descriptive statistics for the entire forum population dataset are shown in Appendix~\ref{sec:appendix_statistics}.

\begin{table}[h]
\small
\caption{Descriptive Statistics of Behavior Indicators for \textit{Top Posters}}
\begin{center}
\label{tables:descriptive_statistics_indicators}
\input{tables/descriptive_statistics_indicators}
\end{center}
\end{table}

\subsubsection{\textbf{Mann-Whitney U Tests}}
We computed Mann-Whitney U tests to assess if the distributions of sub-indicators for the crossover users differed from non-crossover users. A Mann-Whitney U test compares two groups by ranking their respective values from low to high and then comparing the average rank of the two groups. This test was favored over more common parametric tests because all sub-indicators do not follow a normal distribution. The assumptions behind the tests are that the distributions of the data from the two groups are independent; they follow a similar shape and are ordinal or continuous. The sub-indicators all respect these assumptions.

For each sub-indicator, the null hypothesis ($H_o$) was that there is no difference between crossover users and non-crossover users. The alternative hypothesis ($H_a$) was that there is a difference between crossover users and non-crossover users. The significance level of the tests was set to 0.05, meaning that there is a 5\% risk of concluding that a difference exists when there is no difference. 

Below, we report the group’s (crossover or non-crossover) mean, standard deviation, median, \textit{mannwhitneyu} statistics ($U$), and p-value. Also, to measure the effect size, we used the common language effect size introduced by~\cite{mcgraw1992}. It represents the proportion of favorable pairs that support one direction. In this study, the effect size for each sub-indicator represents the proportion of favorable pairs for the group that scored the highest for that sub-indicator (which can be inferred from the mean and/or median).

\subsection{\textbf{Results of Crossover Users Analyses}}
\label{sec:results-crossover-users}
The results of the macro assessment on crossover users are presented below. 


\subsubsection{\textbf{A Minimum of 7\% of Crossover Users}}
\label{sec:correlating-forum-users-with-crime}
A total of  21,726 users had a username of at least five characters. Out of them, 1,557 posted in one of the 38 cybercrime forums between 2015 and 2020, representing 7.2\% of the public forum population. Of the \textit{Top Poster} dataset, a total of 6,433 individuals had a username of at least five characters. Out of them, 510 were crossover users: they posted at least once in one of the 38 cybercrime forums between 2015 and 2020. These crossover users represent 7.9\% of the \textit{Top Poster} dataset. 

\subsubsection{\textbf{Diversified Cybercrime Forums and Limited Involvement}}
We then investigated the branding behind cybercrime forums. Of the 38 forums, seven focused on \textit{hacking}, seven on cracking (cracked software) or leaked information (e.g., lists of usernames and passwords), six on carding (credit card fraud), while three were cryptomarkets (marketplaces hosted on Tor), one involved money laundering discussions, one was specialized in sharing black hat SEO techniques, and thirteen gathered various discussions, sales and questions on various content related to cybercrime (with no clear specific branding). Also, of these 38 cybercrime forums, 17 were hosted on the \textit{clearnet}, meaning that they could be visited via a web browser such as Chrome. The remaining 21 were hosted on The Onion Router (Tor) network, known as the \textit{darknet}. 

Figure 3 illustrates the 38 forums in terms of (i) the forum’s accessibility via the clearnet or the darknet, (ii) the forum’s main branding, (iii) the total number of crossover users who commented on it, and (iv) the total number of comments. This data includes information from all crossover users identified (and not only the \textit{Top Posters}). The figure is a tree map where the size of the boxes represents the number of crossover users in the sample who interacted in the cybercrime forum (specified under the name of the cybercrime forum). The color scale represents the number of comments on each cybercrime forum. Appendix~\ref{sec:appendix_list_cybercrime_forums} shows the complete list and information on all 38 forums.

\begin{figure*}
  \center
    \includegraphics[width=0.7\textwidth]{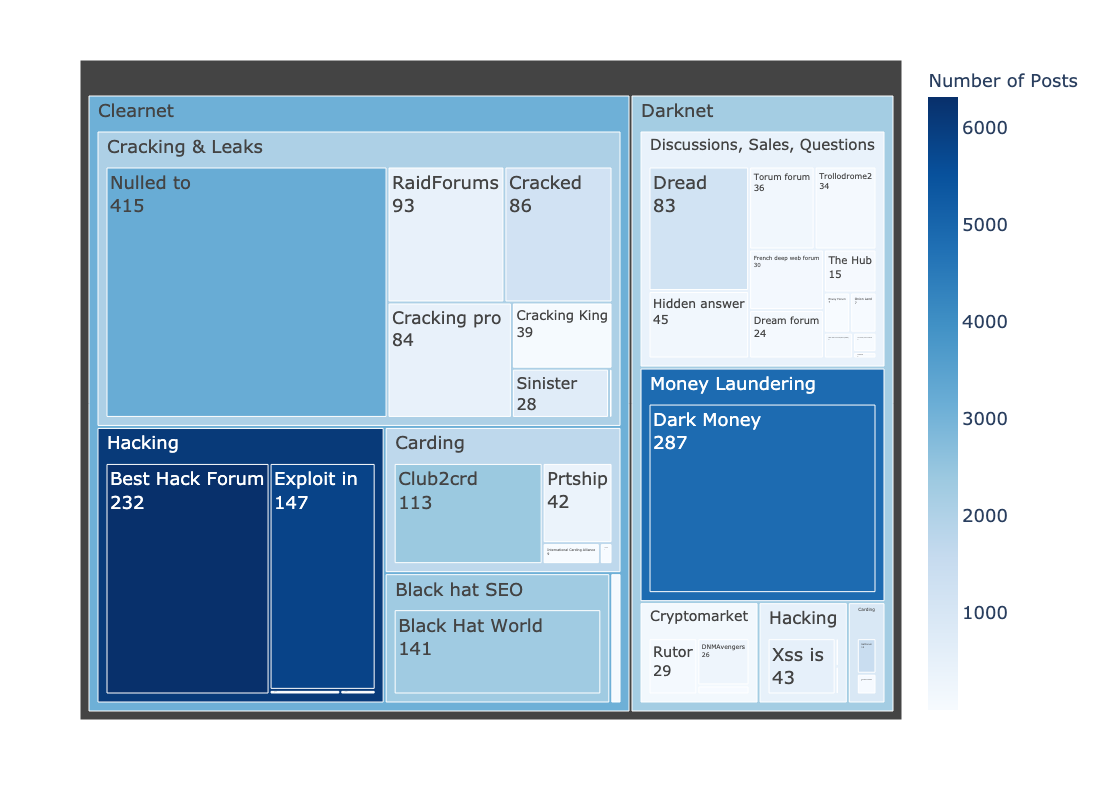}
    \caption{Cybercrime forums in relation to accessibility, main branding,number of comments and number of crossover users}
    \label{fig:distribution-of-dual-actors-in-crime-forums}   
  \center
\end{figure*}

As shown in Figure 3, in terms of number of crossover users, the most popular forums are \textit{Nulled to} (cracking and leaks), \textit{Dark Money} (money laundering), \textit{Best Hack Forum} (hacking), \textit{Exploit In} (hacking), \textit{Black Hat World} (Black Hat SEO), and \textit{Club2crd} (carding). These cybercrime forums are also the ones with the greatest number of comments, although in a different order. Overall, crossover users commented on a variety of cybercrime forums, from cracking (and leaks) to hacking or money laundering. In terms of posting patterns, crossover users favored cybercrime forums hosted on the \textit{clearnet} over those hosted on the \textit{darknet}. That is, 61\% of crossover users commented \textbf{only} on cybercrime forums hosted on the \textit{clearnet}.

Also, their participation in cybercrime forums was \textbf{limited}. Crossover users posted, on average, 21 comments on cybercrime forums (std=75), with a minimum of one and a maximum of 1,383. More importantly, 50\% of cross over users posted three comments and 75\% posted only ten comments!

\subsubsection{\textbf{A Relatively Indistinguishable Group}}
With crossover users identified, we used \textit{the map} generated in the case study section to visualize where these users were positioned in the public forum. As shown in Figure 4, which considers only the \textit{Top Poster} dataset, crossover users are positioned all over the public forum. In other words, these users do not cluster in specific groups of the public forum as identified by the UMAP algorithm.  This finding is further supported by the formal statistical analysis below.

\begin{figure}
  \centering
    \includegraphics[width=0.5\textwidth]{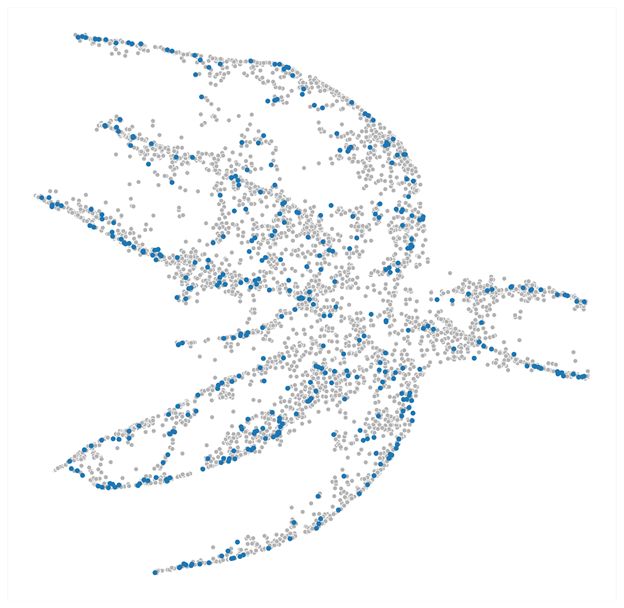}

    \caption{The \textit{Map} with crossover users (blue dots)}
    \label{fig:bird_with_dual_actors}   
  \centering
\end{figure}


Table~\ref{tables:mann_whitney_results} shows the results of the Mann-Whitney U tests for the \textit{Top Poster} dataset. The results for the entire forum population dataset are available in Table~\ref{sec:appendix_statistics} of the Appendix.

Table~\ref{tables:mann_whitney_results} shows that four out of the 13 sub-indicators (30\%) display statistical significance and suggest that there is a difference between crossover users and non-crossover users. However, analysis of the descriptive statistics for these four sub-indicators shows that the absolute differences reported are minimal. These minimal differences are also shown in the small effect sizes, which oscillate around 50\%. 

The same tests computed on the whole population (Appendix~\ref{sec:appendix_statistics}) display eight significant relationships out of 13 sub-indicators (61\%). However, the minimal differences in the descriptive statistics and the small effect sizes both prevent us from concluding that there exist significant differences that differentiate crossover users from non-crossover users. 

This absence of noticeable differences between the two groups is quite informative. Indeed, the fact that almost all indicators are non-significant and that, when there is statistical significance, the effect size is small, suggest that either the two populations are practically indistinguishable given these characteristic variables, or that they are strongly overlapping, hence hinting at a larger crossover user group in the public forum population.

\begin{table*}[h]
    \small
    \caption{Mann-Whitney U Test Results for the \textit{\textbf{Top Poster}} Dataset}
\begin{center}
    \label{tables:mann_whitney_results}
    \input{tables/mann_whitney_results}
\end{center}
\end{table*}

\section{Discussion}\label{sec:discussion}

\textbf{Public Forum Population and Cybercrime Participation Overlap.} This research explores how users of a public forum on internet marketing have ties to cybercrime. The three actors in the case study actively used the public forum as source of information, products and services related to their business which, by 2017 and 2018, had ties to cybercrime activities. Also, according to the forum map, Actors 1 and 2 were positioned in more opportunistic groups with users who spoke less, on average, while Actor 3 was positioned near the center map, in no specific groups. Overall, there was nothing special about them; they are typical users of the forum. Plus, none of them ever mentioned the cybercrime activities they were involved in, which took place between 2017 and 2018. The neutrality of the IT tasks they performed allowed them to conceal the potential maliciousness of their activities in the forum~\cite{leukfeldt2020, bijlenga2018}. The findings of this case study are not unique. We extrapolated the analysis to all forum participants and a lower-bound estimate of 7.2\% of crossover users was found. Further studies should look more deeply into this interplay: how the activities of crossover users in public forums (non branded as cybercrime) relate to activities in cybercrime forums.

Moreover, given that the general description of the forum mentions "find mutually beneficial contacts and partners" and given that the three case study actors used the forum as a market to exchange products and services related to internet marketing, the public forum hosts a market. This market recalls traditional informal markets where the product or service is not necessarily illegal; it is rather the means by which it is produced or distributed that is illegal~\cite{ojo2013, hallerportes2005, castellportes1989}. Thus, this study shows that, since some users do discuss on both spaces, there exists an overlap between informal and criminal online settings, just as in traditional settings\cite{sabet2015informality,shapland2004informal,walle2008matrix}. This was also observed  in previous studies on freelancer platforms~\cite{farooqi2017,motoyama2011,garg2013} which are known to be unregulated online spaces~\cite{schmidt2017, drahokoupil2017, drahokoupil2016}.

\textbf{Hiding in Plain Sight.}
Moreover, the posting behavior of crossover users is relatively indistinguishable from that of non-crossover users in the public platform. Plus, the three case study actors are not part of the crossover user sample although we know they are involved in cybercrime activities. These results suggest that the neutrality of IT tasks~\cite{leukfeldt2020, bijlenga2018} may allow crossover users to behave the same as non-crossover users in informal settings, just like the actors in the case study. It might also suggest that the crossover user sample is a minimum and there exist a higher number of crossover users that were not identified through the cross-correlation method presented above. In such cases, the overlap would be larger than estimated (hence the \textit{lower bound} mention). In both cases, the results point toward a need to investigate further these informal spaces that may represent a hotbed for IT tasks surrounding cybercrime operations. 

Although specialization is known to characterize the cybercrime industry~\cite{lusthaus2018industry, huang2018systematically, Thomas2015}, recent studies have found that specialization, or "as-a-service", advertising in underground forums is quite limited~\cite{Wegberg2018, akyazi2021measuring}. The users in the public forum may be part of the cybercrime specialization trend observed in previous studies~\cite{lusthaus2018industry, huang2018systematically, Thomas2015}. However, the neutrality of their work (e.g., building websites, managing servers, translating texts) could leave a large proportion of them out of cybercrime forums. This could explain why  "as-a-service" offerings are limited in cybercrime forums~\cite{akyazi2021measuring}: other settings ---less targeted by researchers and law enforcement officers--- offer many of these services. The neutrality of the tasks may moreover facilitate the recruitment process, allowing a form of "hiding in plain sight". These findings reinforce the need to study cybercrime participation beyond forums that advertise cybercrime as their branding. Further research should investigate, when and how users in online informal settings end up participating in cybercrime. 

\textbf{Limited Involvement.}
However, 75\% of crossover users posted fewer than 10 comments in any cybercrime forum, suggesting limited participation. They also favored cybercrime forums hosted on the clearnet over those hosted on the darknet. The darknet is mainly linked to the anonymous Tor network, which has a reputation for fostering criminal activities~\cite{owen2015tor, faizan2019exploring}. This suggests that the crossover users studied may limit their cybercrime participation, at least in forums that clearly embody criminal branding. This is in line with Sabet (2015)~\cite{sabet2015informality}, who argued that informal workers from traditional markets preferred to avoid criminal ties when possible. To better understand the reality of informal workers in online settings, there is a need to assess what crossovers do on cybercrime forums and their degree of involvement in them. 

Researching, on a broader scale, the opportunity landscape of these users could further our understanding on cybercrime participation by what appear to be informal workers. This is especially true when considering the three aims of informal economies discussed by Portes and Haller (2005)~\cite{hallerportes2005}: survival, dependent exploitation (such as decreased labor costs) and growth (including capital accumulation, solidarity and flexibility). Informal markets provide jobs to otherwise unemployed individuals and lower costs for products and services, and they foster innovation~\cite{ojo2013, hallerportes2005}. Hence, there is a need to consider the positive effects of such informal markets, and the business landscape they offer, when understanding how and when such workers end up contributing to cybercrime. Furthermore, considering their limited involvement, both focusing on changing their landscape and raising awareness specifically within this population on the harms induced by cybercrime represent alternative approaches to preventing cybercrime participation.


\section{Limits and Future Studies}\label{sec:limits}
There are several limits to the findings of this study, limits which open new research avenues for future studies on the topic, that need to be mentioned. A first limit lies in the cultural origin of the public forum, which mainly includes Russian-speaking individuals. Analyzing whether users from other online informal settings have connections to cybercrime could provide a further understanding of the problem. Also, the identification of crossover users depended on the visibility of cybercrime forums by Flare Systems. Flare Systems monitors over 100 forums, yet the company is not necessarily focused on Russian-speaking forums. To expand the visibility of potential cybercrime forums, partnerships with other organizations could be developed in future studies. 

Moreover, the name filtering approach of keeping only usernames with a minimum of five characters limits the scope of the results. Many alternative approaches could be taken in further studies. For example, individuals registering in different forums with a slightly similar username, such as \textit{Sarik9} and \textit{Sarik10}, could be identified. Such a method would likely yield a higher lower bound. Finally, future studies could improve the posting behavior indicators, using machine learning techniques to analyze comments beyond their categories or subcategories.

\section{Conclusion}\label{sec:conclusion}
This is the first study to formally quantify how users of an internet marketing public forum, a space for informal exchanges, have ties to cybercrime activities. We conclude that crossover users are a substantial part of the population in the public forum, and even though they have been overlooked, their aggregated effect in the ecosystem must be considered. This study opens new research questions on cybercrime participation that should consider online spaces beyond their cybercrime branding. We hope that these findings can be used as a stepping stone for future studies uncovering the territories of online informal markets and their potential ties with cybercrime.

\section*{Acknowledgments}
The authors would like to thank Veronica Valeros from the Stratosphere Laboratory for part of the data gathering process, Avast Software for partial funding of this research, GoSecure Company for partial funding of this research, and Flare Systems for their access to the forum's data. Also, to Anna Shirokova for her help in the translations and the context of the interactions.


\bibliographystyle{plain}
\bibliography{bibliography}

\section*{Appendix}\label{sec:appendix_1}

\appendix

\section{List of Cybercrime Forums }
\label{sec:appendix_list_cybercrime_forums}
Table~\ref{tables:cybercrime_forums_appendix} shows the complete list of 38 cybercrime forums used in this research.

\begin{table*}[!htbp]
\small
\caption{List of the 38 Cybercrime Forums Found}
\begin{center}
\label{tables:cybercrime_forums_appendix}
\input{tables/cybercrime_forums_appendix}
\end{center}
\end{table*}

\section{Descriptive Statistics on all Users of Public Forum}
\label{sec:appendix_statistics}
Table~\ref{tables:descriptive_statistics_indicator_appendix} shows the descriptive statistics for the public forum population dataset.

\begin{table*}[!htbp]
\small
\caption{Descriptive Statistics the Public Forum Population Dataset}
\begin{center}
\label{tables:descriptive_statistics_indicator_appendix}
\input{tables/descriptive_statistics_indicator_appendix}
\end{center}
\end{table*}

\section{Results of the Mann-Whitney U Test for the Whole Population of the Public Forum}
\label{sec:appendix_statistics}
Table~\ref{tables:mann_whitney_results_appendix} shows the Mann-Whitney U tests results for the public forum population dataset.

\begin{table*}[!htbp]
\small
\caption{Mann-Whitney U Tests for the Public Forum Population Dataset}
\begin{center}
\label{tables:mann_whitney_results_appendix}
\input{tables/mann_whitney_results_appendix}
\end{center}
\end{table*}

\end{document}

%% file: tables/Categories.tex
\begin{tabular}{llc}
\multicolumn{1}{c}{\textbf{Category}} & \multicolumn{1}{c}{\textbf{Subcategories}}                                                        & \textbf{\% of Comments} \\
About Monetizing Sites                & \begin{tabular}[c]{@{}l@{}} Partnership Programs, General Questions about \\ Making Money on Sites, YouTube Monetization \end{tabular}    & 20\%                    \\
\hline
Not About Work                        & Meetings and Gatherings, Smoking Room, About the Site and Forum                               & 17\%                    \\
\hline
Site Building                         & \begin{tabular}[c]{@{}l@{}}Domain Names, Hosting and Servers for Websites, Web Analytics,\\ Copywriting \end{tabular}& 16\%                    \\
\hline
Communication of Professionals        & Cryptocurrencies,   Ecommerce, Social Media Marketing                                           & 14\%                    \\
\hline
Practical Optimization Issues         &\begin{tabular}[c]{@{}l@{}} Popular SEO and   SEO Newbie Questions, Doorways and Cloaking,\\ General Optimization Issues \end{tabular}      & 13\%                    \\
\hline
Search Engine                         & Yandex, Site   Directories, Google                                                              & 10\%                    \\
\hline
Exchange and Sales                    & Buying and   Selling Sites, Digital Goods, Programs and Scripts                                 & 5\%                     \\
\hline
Work and Services for Webmasters      &\begin{tabular}[c]{@{}l@{}} Copywriting   Translations, Social Media Marketing Services, \\ Optimization Promotion and Audit \end{tabular} & 3\%                     \\
\hline
About Purchased Traffic for Websites  & \begin{tabular}[c]{@{}l@{}}Teaser and Banner Advertising, Contextual Advertising,\\ Yandex Direct, Google Ads\end{tabular}              & 2\%                    
\end{tabular}

%% file: tables/descriptive_stratistics_dataset_2017_2018.tex
\begin{tabular}{lrrrr}
\multicolumn{2}{l}{All Users. N=23,348} &       &            &     \\
                           & Min                & Max   & Mean (std) & Med \\
N. Comments                & 1                  & 6,603 & 30   (151) & 4   \\
N. threads                 & 1                  & 2,013 & 11   (50)  & 2   \\
N. Category                & 1                  & 9     & 2   (2)    & 1   \\
                           & & & & \\
\multicolumn{2}{l}{Top 30\% Users. N=6,924}     &       &            &     \\
                           & Min                & Max   & Mean (std) & Med \\
N. Comments                & 10                 & 6,603 & 92   (267) & 27  \\
N. threads                 & 1                  & 2,013 & 34   (88)  & 12  \\
N. Category                & 1                  & 9     & 4   (2)    & 3  
\end{tabular}

%% file: tables/three_actors_public_comments.tex
\begin{tabular}{cccc}
\hline
\textbf{} & \textbf{Actor 1} & \textbf{Actor 2} & \textbf{Actor 3}  \\
\hline
    \textbf{N. Comments} & 1,385 & 172 & 471\\
    \textbf{N. Threads} &  759 & 69 & 331\\
    \textbf{Activity Period} & 2009-2020 & 2012-2019 & 2010-2019 \\
\hline
\end{tabular}

%% file: tables/descriptive_statistics_indicators.tex
\begin{tabular}{lrrrrrrrr}
N=6,924                              & \textbf{Min} & \textbf{Max} & \textbf{Mean (std)} & \textbf{Med} \\
\textbf{Activity   Rate}               &              &              &                     &              \\
N. posts             & 10           & 6,603        & 92 (267)            & 27           \\
N. days active       & 1            & 708          & 41 (67)             & 17           \\
\textbf{Diversification}             &              &              &                     &              \\
N. cat                  & 1            & 9            & 4 (2)               & 3            \\
N. sub-cat             & 1            & 71           & 7 (9)               & 4            \\
\textbf{Topics   Discussed}          &              &              &                     &              \\
Search Engines                       & 0            & 1,109        & 9 (38)              & 0            \\
Monetizing Sites               & 0            & 3,010        & 18 (75)             & 1            \\
Practical Opt.       & 0            & 2,965        & 12 (56)             & 1            \\
Comm. of Prof.      & 0            & 2,363        & 13 (75)             & 0            \\
Site Building                       & 0            & 2,873        & 14 (81)             & 1            \\
Exch. and Sales                  & 0            & 689          & 4 (17)              & 0            \\
Purch. Traffic & 0            & 880          & 2 (18)              & 0            \\
Work Webmaster     & 0            & 296          & 3 (11)              & 0            \\
Not About Work                        & 0            & 3,532        & 16 (129)            & 0           
\end{tabular}

%% file: tables/mann_whitney_results.tex
\begin{tabular}{lrrr|rrrr|rr}
                         & \textbf{}     & \multicolumn{2}{l}{\textbf{\begin{tabular}[c]{@{}l@{}}Crossover Users\\    \\ N= 510\end{tabular}}} & \textbf{}     & \multicolumn{2}{l}{\textbf{\begin{tabular}[c]{@{}l@{}}Non-Crossover Users\\    \\ N=6,414\end{tabular}}} & \multicolumn{3}{l}{\textbf{Statistics}}                 \\
                         & \textbf{Mean} & \textbf{Std}                                & \textbf{Med}                                & \textbf{Mean} & \textbf{std}                                   & \textbf{Med}                                  & \textbf{Mannwhitneyu} & \textbf{p-value} & \textbf{Effect Size} \\
\textbf{Activity Rate}   &               &                                             &                                                &               &                                                &                                                  &                       &                  &              \\
N. posts                 & 99.79         & 289.63                                      & 29.0                                           & 91.10         & 264.88                                         & 27.0                                             & 1,559,642               & 0.04             & 0.52         \\
N. days active           & 44.68         & 68.89                                       & 19.0                                           & 40.36         & 67.16                                          & 17.0                                             & 1,526,452               & 0.01             & 0.53         \\
\textbf{Diversification} &               &                                             &                                                &               &                                                &                                                  &                       &                  &              \\
N. cat.                  & 3.69          & 2.33                                        & 3.0                                            & 3.53          & 2.25                                           & 3.0                                              & 1,577,524               & 0.09             & 0.52         \\
N. sub-cat               & 7.70          & 9.23                                        & 5.0                                            & 7.19          & 8.47                                           & 4.0                                              & 1,598,917               & 0.20             & 0.51         \\
\textbf{Activity in Categories} &               &                                             &                                                &               &                                                &                                                  &                       &                  &              \\
Search Engine            & 8.64          & 30.23                                       & 0.0                                            & 9.29          & 38.61                                          & 0.0                                              & 1,619,719               & 0.34             & 0.50         \\
Monetizing Sites         & 21.73         & 139.65                                      & 1.0                                            & 17.80         & 67.07                                          & 1.0                                              & 1,629,106               & 0.44             & 0.50         \\
Practical Opt.           & 10.81         & 32.21                                       & 1.0                                            & 11.73         & 57.74                                          & 1.0                                              & 1,595,856               & 0.17             & 0.51         \\
Comm. of Prof.           & 19.06         & 86.82                                       & 0.0                                            & 12.47         & 74.21                                          & 0.0                                              & 1,515,778               & 0.00             & 0.54         \\
Site Building            & 13.28         & 45.15                                       & 1.0                                            & 14.52         & 83.63                                          & 1.0                                              & 1,579,436               & 0.08             & 0.52         \\
Exch. and Sale           & 4.18          & 13.72                                       & 0.0                                            & 4.21          & 17.37                                          & 0.0                                              & 1,587,799               & 0.10             & 0.51         \\
Purch. Traffic           & 2.54          & 18.54                                       & 0.0                                            & 1.92          & 17.90                                          & 0.0                                              & 1,626,877               & 0.39             & 0.50         \\
Work Webmasters          & 2.45          & 8.36                                        & 0.0                                            & 2.72          & 10.86                                          & 0.0                                              & 1,615,643               & 0.28             & 0.51         \\
Not About Work           & 17.11         & 102.36                                      & 0.0                                            & 16.43         & 130.75                                         & 0.0                                              & 1,561,045               & 0.02             & 0.52        
\end{tabular}

%% file: tables/cybercrime_forums_appendix.tex
\begin{tabular}{lllll}
\textbf{Forum Name}                              & \textbf{Hosted} & \textbf{Type}                 & \textbf{N. Actors} & \textbf{N. Posts} \\
\textbf{Nulled to}                      & Clearnet        & Cracking and Leaks            & 415                & 3205                 \\
\textbf{Dark Money}                     & Darknet         & Money Laundering              & 287                & 4872                 \\
\textbf{Best Hack Forum}                & Clearnet        & Hacking                       & 232                & 6311                 \\
\textbf{Exploit in}                     & Clearnet        & Hacking                       & 147                & 5845                 \\
\textbf{Black Hat World}                & Clearnet        & Black hat SEO                 & 141                & 2319                 \\
\textbf{Club2crd}                       & Clearnet        & Carding                       & 113                & 2407                 \\
\textbf{RaidForums}                     & Clearnet        & Cracking and Leaks            & 93                 & 488                  \\
\textbf{Cracked}                        & Clearnet        & Cracking and Leaks            & 86                 & 1212                 \\
\textbf{Cracking pro}                   & Clearnet        & Cracking and Leaks            & 84                 & 482                  \\
\textbf{Dread}                          & Darknet         & Discussions, Sales, Questions & 83                 & 1207                 \\
\textbf{Hidden answer}                  & Darknet         & Discussions, Sales, Questions & 45                 & 187                  \\
\textbf{Xss is}                         & Darknet         & Hacking                       & 43                 & 538                  \\
\textbf{Prtship}                        & Clearnet        & Carding                       & 42                 & 383                  \\
\textbf{Cracking King}                  & Clearnet        & Cracking and Leaks            & 39                 & 73                   \\
\textbf{Torum forum}                    & Darknet         & Discussions, Sales, Questions & 36                 & 202                  \\
\textbf{Trollodrome2}                   & Darknet         & Discussions, Sales, Questions & 34                 & 90                   \\
\textbf{French deep web forum}          & Darknet         & Discussions, Sales, Questions & 30                 & 138                  \\
\textbf{Rutor}                          & Darknet         & Cryptomarket                  & 29                 & 93                   \\
\textbf{Sinister}                       & Clearnet        & Cracking and Leaks            & 28                 & 738                  \\
\textbf{DNMAvengers}                    & Darknet         & Cryptomarket                  & 26                 & 280                  \\
\textbf{Dream forum}                    & Darknet         & Discussions, Sales, Questions & 24                 & 94                   \\
\textbf{The Hub}                        & Darknet         & Discussions, Sales, Questions & 15                 & 78                   \\
\textbf{SatForum}                       & Darknet         & Carding                       & 12                 & 1480                 \\
\textbf{International Carding Alliance} & Clearnet        & Carding                       & 9                  & 15                   \\
\textbf{Deutschland}                    & Clearnet        & Discussions, Sales, Questions & 7                  & 66                   \\
\textbf{Verified Carders}               & Darknet         & Carding                       & 7                  & 32                   \\
\textbf{Envoy Forum}                    & Darknet         & Discussions, Sales, Questions & 7                  & 13                   \\
\textbf{Onion Land}                     & Darknet         & Discussions, Sales, Questions & 7                  & 44                   \\
\textbf{Wall Street forum}              & Darknet         & Cryptomarket                  & 5                  & 27                   \\
\textbf{Dark Anti French System (DFAS)} & Darknet         & Discussions, Sales, Questions & 5                  & 104                  \\
\textbf{Criminality French Market}      & Darknet         & Discussions, Sales, Questions & 3                  & 12                   \\
\textbf{Xaker26}                        & Clearnet        & Hacking                       & 2                  & 46                   \\
\textbf{CardVilla}                      & Clearnet        & Carding                       & 2                  & 2                    \\
\textbf{Sinfulsite}                     & Clearnet        & Cracking and Leaks            & 1                  & 2                    \\
\textbf{Hermes}                         & Darknet         & Discussions, Sales, Questions & 1                  & 1                    \\
\textbf{Main Helium}                    & Darknet         & Hacking                       & 1                  & 5                    \\
\textbf{CryptBB}                        & Darknet         & Hacking                       & 1                  & 5                    \\
\textbf{GreySec}                        & Clearnet        & Hacking                       & 1                  & 1                   
\end{tabular}

%% file: tables/descriptive_statistics_indicator_appendix.tex
\begin{tabular}{lrrrrrrrr}

N=23,348                             & \textbf{Min} & \textbf{Max} & \textbf{Mean (std)} & \textbf{Med}  \\

\textbf{Activity   Rate}                      &                    &                      &                           &                   \\ 
N, posts                      & 1                  & 6,603                & 29 (151)                  & 4                 \\ 
N. days active                & 1                  & 708                  & 14 (41)                   & 3                 \\ 
\textbf{Diversification}                      &                    &                      &                           &                   \\ 
N. cat                 & 1                  & 9                    & 2 (2)                     & 1                 \\ 
N. sub-cat             & 1                  & 71                   & 3 (5)                     & 1                 \\
\textbf{Topics   Discussed}                   &                    &                      &                           &                  \\
Search Engines                       & 0                  & 1,109                & 3 (21)                    & 0                 \\
Monetizing Sites               & 0                  & 3,010                & 6 (42)                    & 0                 \\ 
Practical Opt.        & 0                  & 2,965                & 4 (31)                    & 0                 \\ 
Comm. of Prof.       & 0                  & 2,363                & 4 (41)                    & 0                 \\ 
Site Building                        & 0                  & 2,873                & 5 (45)                    & 0                 \\ 
Exch. and Sales                   & 0                  & 689                  & 2 (10)                    & 0                 \\ 
Purchased Traffic & 0                  & 880                  & 1 (10)                    & 0                 \\ 
Work Webmasters     & 0                  & 296                  & 1 (6)                     & 0                 \\ 
Not About Work                       & 0                  & 3,532                & 5 (71)                    & 0                 \\ 
\end{tabular}

%% file: tables/mann_whitney_results_appendix.tex
\begin{tabular}{llll|llll|ll}
                         &       & \multicolumn{2}{l}{\begin{tabular}[c]{@{}l@{}}Crossover Users\\    \\ N= 1,557\end{tabular}} &       & \multicolumn{2}{l}{\begin{tabular}[c]{@{}l@{}} Non-Crossover Users\\    \\ N=21,791\end{tabular}} & \multicolumn{3}{l}{Statistics} \\
                         & Mean  & Std                                       & Med                                    & Mean  & std                                         & Med                                     & Mannwhitneyu  & p-value & Effect Size  \\
\textbf{Activity Rate}   &       &                                           &                                           &       &                                             &                                             &               &         &      \\
N. posts                 & 34.76 & 171.77                                    & 4.0                                       & 28.87 & 149.23                                      & 4.0                                         & 15,875,519      & 0.0000  & 0.53 \\
N. days active           & 16.24 & 44.15                                     & 3.0                                       & 13.44 & 40.40                                       & 2.0                                         & 15,666,700      & 0.0000  & 0.54 \\
\textbf{Diversification} &       &                                           &                                           &       &                                             &                                             &               &         &      \\
N. cat.                  & 2.13  & 1.81                                      & 1.0                                       & 1.97  & 1.68                                        & 1.0                                         & 16,007,108      & 0.0000  & 0.53 \\
N. sub-cat               & 3.55  & 6.07                                      & 1.0                                       & 3.14  & 5.34                                        & 1.0                                         & 16,029,862      & 0.0000  & 0.53 \\
\textbf{Activity in Categories} &       &                                           &                                           &       &                                             &                                             &               &         &      \\
Search Engine            & 3.00  & 17.75                                     & 0.0                                       & 2.94  & 21.36                                       & 0.0                                         & 16,895,351      & 0.3573  & 0.50 \\
Monetizing sites         & 7.50  & 80.50                                     & 0.0                                       & 5.65  & 37.24                                       & 0.0                                         & 16,849,367      & 0.2950  & 0.50 \\
Practical opt.           & 3.82  & 19.08                                     & 0.0                                       & 3.81  & 31.76                                       & 0.0                                         & 16,959,295      & 0.4905  & 0.50 \\
Comm. of prof.           & 6.52  & 50.43                                     & 0.0                                       & 3.89  & 40.65                                       & 0.0                                         & 16,104,023      & 0.0000  & 0.53 \\
Site building            & 4.75  & 26.53                                     & 0.0                                       & 4.62  & 45.83                                       & 0.0                                         & 16,227,874      & 0.0002  & 0.52 \\
Exch. and sale           & 1.61  & 8.10                                      & 0.0                                       & 1.45  & 9.62                                        & 0.0                                         & 16,307,064      & 0.0001  & 0.52 \\
Purch. traffic           & 0.93  & 10.68                                     & 0.0                                       & 0.66  & 9.76                                        & 0.0                                         & 16,832,449      & 0.1637  & 0.50 \\
Work Webmasters          & 0.97  & 4.94                                      & 0.0                                       & 0.96  & 6.04                                        & 0.0                                         & 16,778,840      & 0.1291  & 0.50 \\
Not about work           & 5.66  & 59.09                                     & 0.0                                       & 4.89  & 71.32                                       & 0.0                                         & 16,548,104      & 0.0020  & 0.51
\end{tabular}